\documentclass{aa}[8pt]

\usepackage{graphicx}
\usepackage{color}
\usepackage{amsbsy}
\usepackage{mathrsfs}
\usepackage{mathpazo,bm}
\usepackage{amsmath}
\usepackage{multirow}

\newcommand{\ptderiv}[1]{\frac{\partial #1}{\partial t}}

\newcommand{\Ma}{\mathrm{Ma}}
\newcommand{\Rey}{\mathrm{Re}}
\newcommand{\Kn}{\mathrm{Kn}}
\newcommand{\ttimes}[1]{10^{#1}}
\newcommand{\xtimes}[2]{#1 \times 10^{#2}}

\newcommand{\vt}[1]{\mathbf{#1}}       
\renewcommand{\v}[1]{{\boldsymbol #1}} 

\newcommand{\del}{\v{\nabla}}
\newcommand{\grad}{\del}
\newcommand{\Div}{\v{\nabla}\cdot}

\newcommand{\Laplace}{\nabla^2}

\renewcommand{\sun}{\odot}
\newcommand{\Eps}{{\rm Eps}}
\newcommand{\Stk}{{\rm Stk}}
\newcommand{\dv}{\Delta{\v{v}}}
\newcommand{\mearth}{\,$M_{\oplus}$}

\newcommand{\mps}{m\,s$^{-1}$}
\newcommand{\vrms}{$v_{\rm rms}$} 
\newcommand{\vmax}{$v_{\rm max}$} 
\newcommand{\vesc}{$v_{\rm esc}$} 
\newcommand{\Sun}{\odot}

\newcommand{\Eq}[1]{Eq. (\ref{#1})}

\newcommand{\Fig}[1]{Fig.~\ref{#1}}
\newcommand{\fig}[1]{\Fig{#1}}
\newcommand{\Figure}[1]{Figure ~\ref{#1}}

\definecolor{brown}{rgb}{0.42,0.24,0.07}
\definecolor{darkgreen}{rgb}{0.0,0.6,0.00}
\definecolor{purple}{rgb}{0.7,0.0,0.7}

\begin{document}

\title{Embryos grown in the dead zone}
\subtitle{Assembling the first protoplanetary cores\\
in low mass self-gravitating circumstellar disks of gas and solids}

\author{
W. Lyra\inst{1},
A. Johansen\inst{2},
H. Klahr\inst{3}, \&
N. Piskunov\inst{1}
}

\offprints{wlyra@astro.uu.se}

\institute{Department of Physics and Astronomy, Uppsala Astronomical 
Observatory, Box 515, 751\,20 Uppsala, Sweden
\and Leiden Observatory, Leiden University, PO Box 9513, 2300 RA Leiden, The Netherlands
\and Max-Planck-Institut f\"ur Astronomie, K\"onigstuhl 17, 69117 
Heidelberg, Germany}

\date{Received ; Accepted}

\authorrunning{Lyra et al.}
\titlerunning{Embryos grown in the dead zone}

\abstract
{In the borders of the dead zones of protoplanetary disks, the inflow of gas 
produces a local density maximum that triggers the Rossby wave instability. 
The vortices that form are efficient in trapping solids.}
{We aim to assess the possibility of gravitational collapse of the solids 
within the Rossby vortices.} 
{We perform global simulations of the dynamics of gas and solids in a 
low mass non-magnetized self-gravitating thin protoplanetary disk with the 
Pencil Code. We use multiple particle species of radius 1, 10, 30, and 100\,cm. 
The dead zone is modelled as a region of low viscosity.}
{The Rossby vortices excited in the edges of the dead zone are efficient 
particle traps. Within 5 orbits after their appearance, the solids achieve 
critical density and undergo gravitational collapse into Mars sized objects. 
The velocity dispersions are of the order of 10 \mps~  for newly formed 
embryos, later lowering to less than 1\,\mps~ by drag force cooling. After 
200 orbits, 38 gravitationally bound embryos were formed inside the vortices, 
half of them being more massive than Mars.}
{}


\maketitle



\section{Introduction}

The formation of planets is one of the major unsolved problems 
in modern astrophysics. In the standard core accretion 
scenario, sub-$\mu$m grains assemble into progressively 
larger bodies through electrostatic interactions (Natta et al. 2007), eventually growing into 
centimeter and meter sized boulders. Growth beyond this size, 
however, is halted since these boulders have very poor sticking properties 
and are easily destroyed by collisions at the velocities assumed to be 
prevalent 
in circumstellar disks (Benz 2000). Furthermore, 
centimeter and meter sized solids are loosely decoupled 
from the gas, but remain sufficiently small to
be affected by significant gas drag. The resulting
headwind from the sub-Keplerian gas reduces their angular
momentum and forces them into spiral
trajectories onto the star in timescales as short as a
few thousand years (Weidenschilling 1977a). 

A mechanism for overcoming these barriers was presented
by Kretke \& Lin (2007). In the presence 
of sufficient ionization, the gaseous disk couples with the 
ambient weak magnetic field, which 
triggers the growth of the magneto-rotational 
instability (MRI; Balbus \& Hawley 1991). In its 
saturated state, a vigorous turbulence drives accretion onto 
the star by means of magnetic and 
kinetic stresses. However, in the water condensation front (snowline) the
abundant presence of snowflakes effectively removes free 
electrons from the gas, lowering the degree
of ionization. The turbulence is weakened locally and the accretion flow 
is stalled. As the radial 
inflow proceeds from the outer disk, gas accumulates at the 
snowline. Since embedded solid bodies move towards gas pressure maxima
(Haghighipour \& Boss 2003), 
the snowline environment proposed by Kretke \& Lin (2007) is potentially 
an efficient particle trap. This scenario was further explored by 
Brauer et al. (2008), who demonstrated that as solids 
concentrate at this local pressure maximum,  
rapid growth into kilometer sized planetesimals occurs by coagulation.

Kretke et al. (2008) emphasized that an identical mechanism is 
supposed to occur elsewhere in the disk. Ionization 
ought to be present in 
the very inner disk due to the high temperatures, as well 
as in the outer regions where the gas is sufficiently thin for 
cosmic rays 
to penetrate to the disk midplane and provide ionization 
throughout. In between, however, temperatures are too 
low and gas columns too thick to allow sufficient ionization either by 
collisions or by cosmic rays. In the midplane of this region, the gas is 
neutral and the turbulence is largely suppressed (Gammie 1996). As in 
the snowline, the accretion flow from the MRI-active regions  
halts at the borders of this ``dead'' zone, 
where the gas then accumulates.

These models have been tested only by one-dimensional simulations, 
and these tests have therefore not benefitted from an interesting 
development. As shown by Varni\`ere \& Tagger (2006), 
the density pileup at the border 
of the dead zone triggers the Rossby wave instability 
(RWI; Li et al. 2001). The azimuthal symmetry of the problem is broken 
and long-lived anticyclonic vortices are formed as the waves 
break and coalesce. Such 
entities are of significant interest because, by 
rotating clockwise 
in the global counterclockwise Keplerian flow, they amplify the 
local shear and induce a net force on solid particles 
towards their center (Barge \& Sommeria 1995). As shown by 
Klahr \& Bodenheimer (2006), the accumulation of solids under 
these circumstances is likely to lead to high densities. Inaba \& Barge (2006) 
continued the study of Varni\`ere \& Tagger (2006) by including 
solids and confirming that the Rossby vortices excited in the 
borders of the dead zone act as powerful traps, enhancing the local
solids-to-gas ratio by at least an order of magnitude. Unfortunately, 
they used a fluid approach - which limited the maximum particle size 
they could consider - and they did not include the self-gravity 
of the solids, which is crucial to follow the gravitational collapse. 
Studies with interacting particles in the literature include 
a MRI-unstable local box (Johansen et al. 2007) capable of 
producing dwarf planets out of meter sized boulders, a global 
massive disk unstable to gas self-gravity
(Rice et al. 2006) in which concentrations of 0.5 $M_{\oplus}$ are 
seen in the spiral arms, and a simulation that produces 10-100 km 
"sandpile" clumps formed out of mm-sized particles (Cuzzi et al. 
2008).

In this Letter, we build on the studies of Varni\`ere \& Tagger (2006) 
and Inaba \& Barge (2006) by including self-gravitating centimeter 
and meter sized Lagrangian particles to model the solid phase. We 
show that in the vortices launched by the RWI in the borders of the 
dead zone, the solids quickly achieve critical densities and 
undergo gravitational collapse into protoplanetary embryos in the 
mass range 0.1-0.6\mearth. 

\section{Model}

We work in the thin disk approximation, using the vertically averaged 
equations of hydrodynamics. The gas drag is implemented in the same way as 
Paardekooper (2007), interpolating between Epstein and Stokes drag 
(see
online supplement). The back 
reaction of the drag force onto the gas is present.  
The Poisson solver for the particles is a particle-mesh solver based on multiple 
Fourier transforms in a Cartesian grid, as used by Johansen et al. (2007). 

We follow the Varni\`{e}re \& Tagger (2006) dead zone model, which consists of jumps 
in the viscosity profile. We artificially place the inner and outer edges of the dead 
zone at 0.6 and 1.2 times the semi-major axis of Jupiter (5.2\,AU), using Heaviside functions 
to jump from $\alpha=\ttimes{-2}$ to zero inside the dead zone. The parameter $\alpha$ is the 
usual alpha viscosity (Shakura \& Sunyaev 1973). 

The simulations were performed with the {\sc Pencil Code}{\footnote{See http://www.nordita.org/software/pencil-code}} in Cartesian geometry. The resolution was 256$\times$256. The surface density profile followed a 
power law of $\varSigma(r)\propto r^{-1/2}$, and we chose a disk about twice 
as massive as the Minimum Mass Solar Nebula, with surface 
density $\varSigma_0$=300 
g\,cm$^{\rm -2}$ at 5.2 AU. The 
sound speed followed the local isothermal approximation with a 
radial temperature profile $T(r)\propto r^{-1}$. The disk aspect 
ratio was $h$=0.05. For the 
solids, we used $\ttimes{5}$ Lagrangian superparticles and the 
interstellar value for the solids-to-gas ratio ($\ttimes{-2}$). Each superparticle 
therefore contained $\ttimes{-9} M_\sun\simeq\xtimes{2}{24}$ g of 
material. We used multiple particles species, of 1, 10, 30, and 100\,cm radii, 
each represented by 1/4 of the total number of particles. We quote 
time as orbital periods at 5.2\,AU.

\begin{figure*}
  \begin{center}
      \resizebox{.90\textwidth}{!}{\includegraphics{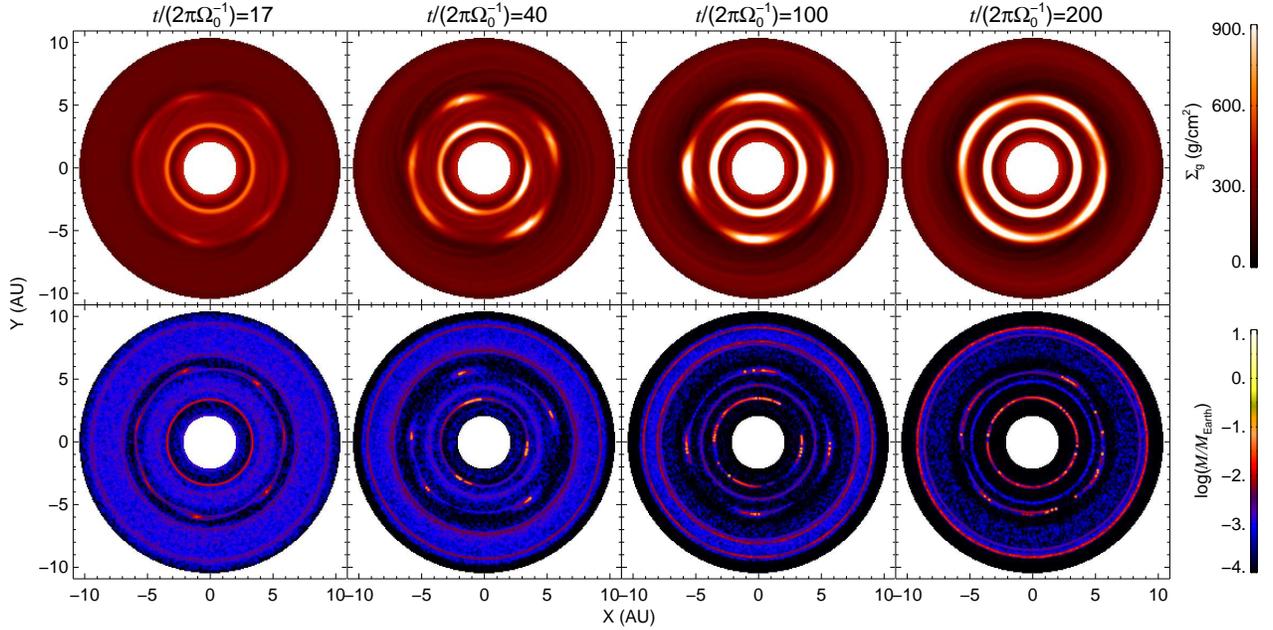}}
\end{center}
\caption[]{The appearance of the disk in the gas (upper panels) and solid (lower panels) phases in selected 
snapshots. The Rossby vortices first appear at 15 orbits. Collapse of the particles into a 
gravitationally bound planetary embryo the size of Mars occurs 5 orbits later. The vortices tend to merge and 
decrease in number, retaining an $m$=4 dominant mode until the end of the simulation, up to which over a dozen 
embryos were formed.} 
\label{fig:deadzone}
\end{figure*}

\begin{figure*}
\begin{center}
  \resizebox{5.6cm}{!}{\includegraphics{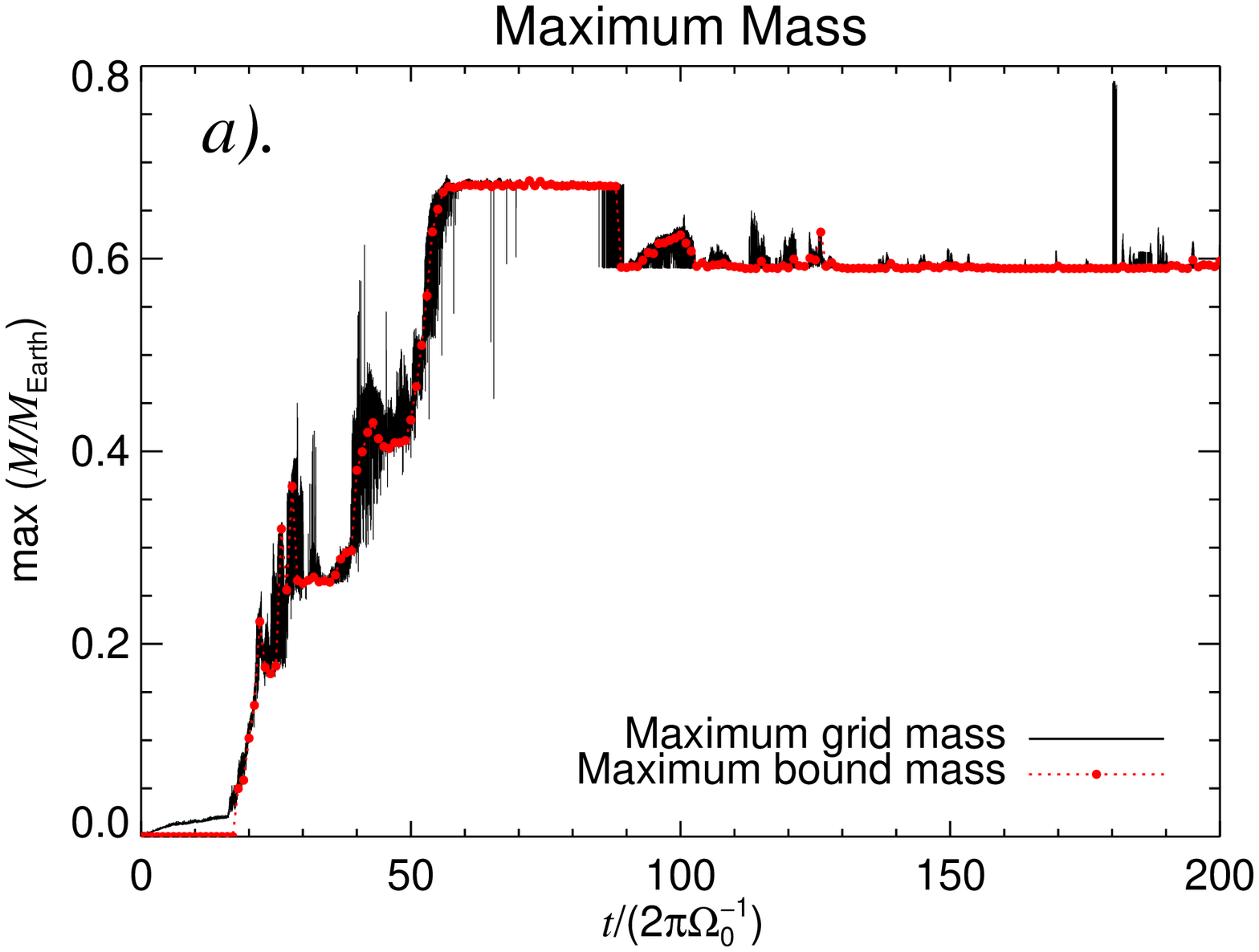}}
\hspace{4mm}
  \resizebox{5.6cm}{!}{\includegraphics{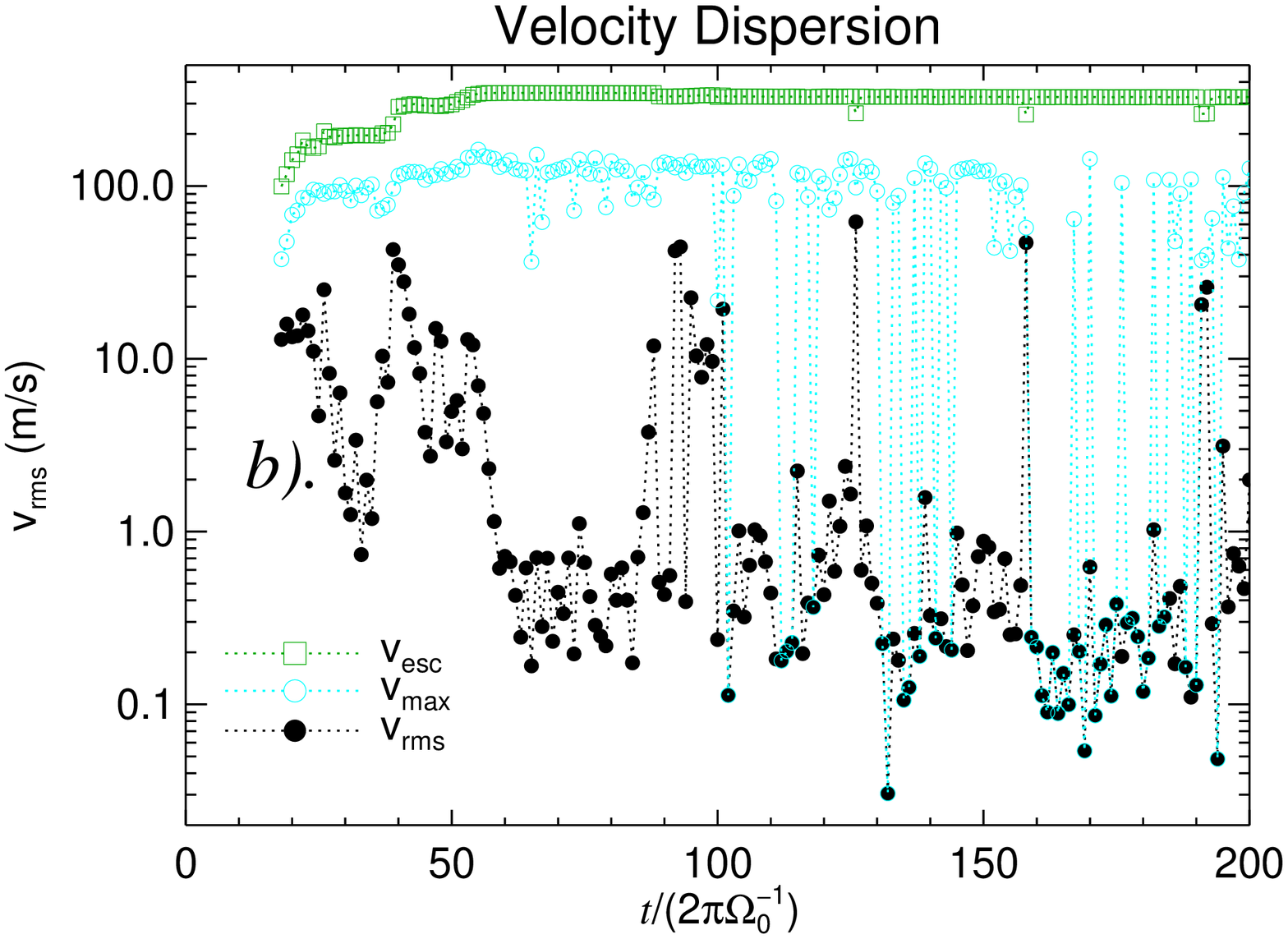}}
\hspace{4mm}
  \resizebox{5.6cm}{!}{\includegraphics{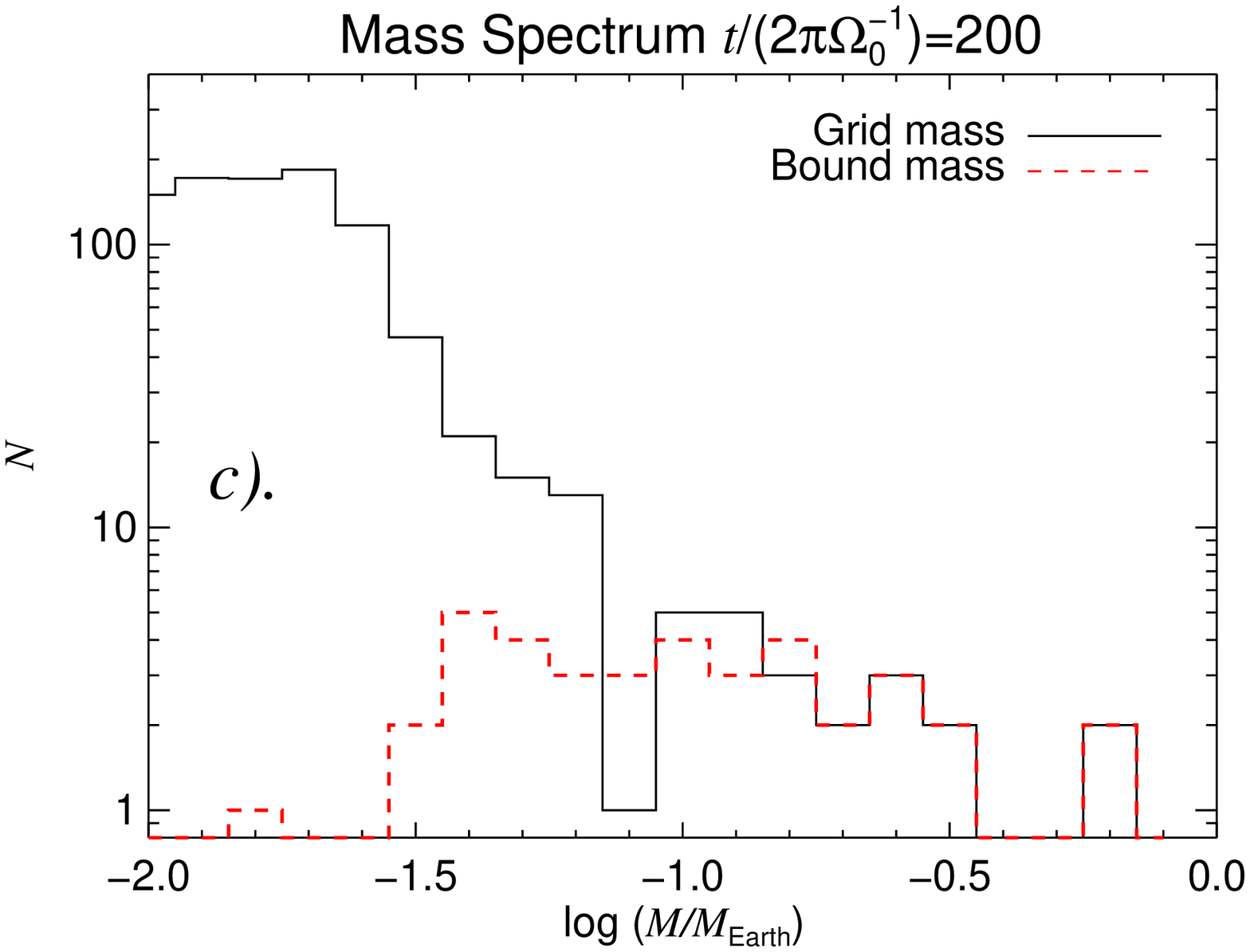}}
\end{center}
\caption[]{{\it a).} Time evolution of the maximum mass of solids. The black solid line is the maximum mass of solids
inside a single cell, and the red dot-dashed line marks the mass that is gravitationally bound. The maximum 
mass settles at 0.59\mearth~ from 90 orbits onwards.\\
           {\it b).} The internal velocity dispersion \vrms~ of the most massive embryo, compared with its 
escaped velocity \vesc. The embryo is so firmly bound that even the maximum internal speed \vmax~ is lower 
than the escape velocity.
Throughout 
most of the simulation, the velocity dispersion is below 1 \mps.\\ 
           {\it c).} The mass spectrum by the end of the simulation. The line
           and color style 
is the same as in Fig.~2a. Over a dozen gravitationally bound embryos in the mass range 0.1-0.6 \mearth~ 
were formed in the vortices launched at the edges of the dead zone.}
\label{fig:max-and-spectrum}
\end{figure*}

\section{Results}

In \Fig{fig:deadzone}, we show snapshots of the appearance of the disk 
for the gas and solids phases.
The vortices triggered by the Rossby wave instability are visible as early as 15 orbits. 
As seen in the solids phase, the particles are trapped 
by the vortical motion and soon reach extremely high densities. After 17 orbits, 
seven vortices appear at the outer edge. After 45 orbits have elapsed, the $m$=4 
mode begins to dominate, persisting until the end of the simulation at 200 orbits, 
their gas surface density peaking at 4.5 times their initial value. In the inner 
edge of the dead zone, at 40 orbits we see a conspicuous $m$=3 mode. By the end 
of the simulation, their surface density has increased by a factor of 8 relative to 
the initial condition, and a weak $m$=2 mode is visible, albeit with far less 
contrast than in the outer disk. 

In \fig{fig:max-and-spectrum}a, we plot the time evolution of the maximum 
concentration of solids. The solid line represents the maximum 
mass of solids contained in a single cell. The red dashed line marks 
the maximum mass that is gravitationally bound. We decide whether boundness 
is present based 
on two criteria. First, we consider the mass inside the Hill's sphere of the clump 
defined by the black line. Particles inside/outside the Hill's sphere are 
added/removed from the total mass, and the Hill's radius recomputed. The 
process is iterated until convergence. This positional criterion is followed by 
a dynamical one. We calculate the velocity dispersion \vrms~ of the particles inside the 
Hill's radius, and compare its value with the escape velocity of the enclosed mass. If \vrms $<$ \vesc, 
we consider that the cluster of particles is bound. We plot the velocity dispersion and escape velocity of the most massive clump in 
\fig{fig:max-and-spectrum}b. The first bound clumps appeared at 18 orbits, with 
masses of 0.050 and 0.036 \mearth. At 20 orbits, four clumps of 0.1\mearth~ are present. 
The mass is that of Mars, constituting a protoplanetary embryo. The efficiency of the vortex trapping mechanism can be more clearly 
appreciated if we consider the time elapsed between the rise of the Rossby 
vortices and the collapse of the trapped particles into a Mars sized object: only 5 orbits. 

Two orbits later, the maximum mass increased to 0.22 \mearth. Nine other clumps 
collapsed into embryos as well, five of them of mass above that of Mars. The maximum bound 
mass reached 0.67\mearth, but settled at 0.59\mearth~ from 89 orbits until 
the end of the simulation at 200 orbits. We observe evidence that the mass 
loss episodes are due to tides from the gas, since the vortices concentrate 
sufficient gas to provide a considerable gravitational pull. In addition to tides, 
erosion (Cuzzi et al. 2008) also plays a role in disrupting clumps of 
smaller particles. We discuss this point further
in the online supplement.
 
\Figure{fig:max-and-spectrum}b also indicates that the velocity dispersion remained 
below 1 \mps~ for most of the simulation. This is of extreme importance because it 
implies that particle encounters are 
gentle enough for destructive collisions to be avoided. The opposite was reported for 
the massive disk models of Rice et al. (2006), where particle 
encounters in the spiral arms occurred at velocities comparable with 
the sound speed. In \fig{fig:max-and-spectrum}b, we plot 
the maximum speed \vmax \ for comparison. It is evident that even \vmax \ remains lower than the 
escape velocity. This indicates that even if destructive collisions occur, the
fragments will remain bound, although the strong drag force felt by the
fragments might delay any gravitational collapse.

We plot the mass spectrum of the formed embryos in \fig{fig:max-and-spectrum}c. The solid black 
line represents the mass of solids inside a cell of the simulation box, without information on boundness. We overplot the 
distribution of bound clumps with the red dashed line. Thirty-eight bound embryos are formed by the end of the 
simulation, twenty of these being more massive than Mars. The two most massive embryos have 0.59 and 0.57 \mearth, and 
are located in the inner and outer edge, respectively. 

\begin{figure}
\begin{center}
  \resizebox{9cm}{!}{\includegraphics{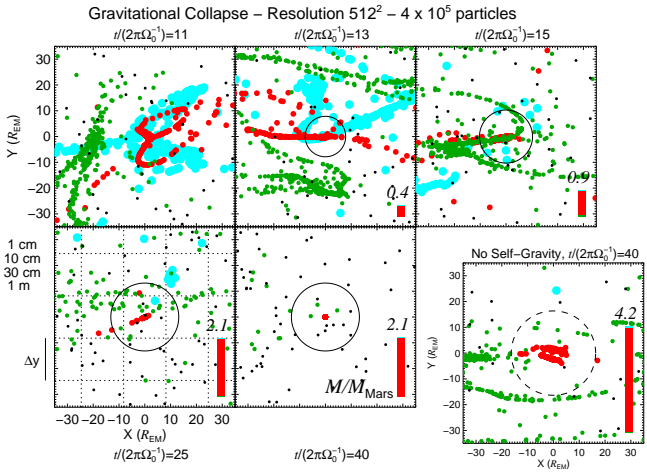}}
\end{center}
\caption[]{Gravitational collapse of the particles into an embryo twice 
the mass of Mars. The unit of length is 
the Earth-Moon mean separation ($R_{\rm EM}$). The lower left panel 
has 
the grid scale overplotted. The circle marks the Hill's radius. A control 
run without selfgravity attains higher cell concentration due to the absence 
of gas tides. The bar shows the mass of the embryo in units of 
$M_{\rm Mars}$ and the fractional mass of each particle species.}
\label{fig:followcollapse}
\end{figure}

A common trait of these embryos is that they consist of an overwhelming 
majority of particles of a single species. For instance, over 98\% of the 
most massive embryo consists of 30\,cm particles. Almost 97\% of the second 
most massive one were 10\,cm particles. In one embryo in the Mars mass bin, 
94\% of the particles were of 1\,cm radius. These different particle radii 
reflect the variations in gas drag strengths experienced by particles of 
different size. Since particles of a given size drift in a similar way, 
their spatial convergence becomes easier than the convergence of particles of
different sizes. The almost single-phasing of the size spectrum of the bound 
clumps also contributes to their low velocity dispersions. Since particles of 
different sizes experience different gas drags, their relative 
velocities are often far higher than the relative velocities between particles 
of identical size 
(Th\'ebault et al. 2008). We indeed see that clumps with a more non-uniform 
distribution of particle sizes usually 
show larger velocity dispersions ($\approx$10 instead of $\approx$1 \mps). 
It is an observed fact that planetesimals are formed 
of similarly sized particles (Scott \& Krot 2005). Although the constituent 
particles appear to be sub-mm grains, different nebula parameters could 
enable smaller grains to be trapped.
 
At higher resolution (512${\rm ^2}$, and $\xtimes{4}{5}$ particles), 
more bound clumps are formed, which extend to lower masses. More 
clumps containing similar quantities of particles of different sizes 
are observed, although the majority of clumps have nearly single phasing. 
The situation does not 
alter significantly when the number of particles is increased to $\ttimes{6}$. 
In Fig.~3, we follow the collapse of one of these clumps at
higher resolution. Although particles of different size are present inside 
the Hill's sphere during the first stages of collapse, most are 
expelled and the collapsed embryo consists primarily of 30\,cm particles. 
A control run without self-gravity achieves higher cell concentration, 
due to the absence of gas tides. At 
later times, the clumps also tend to have low rms speeds and small rms 
radii ($\approx$10$^{\rm 4}$ km), due to the efficient dynamical 
cooling provided by the drag force. Size segregation due to aerodynamical 
sorting is also observed in the control run. Inaba \& Barge (2006) reported 
destruction of the vortices by the drag force backreaction. We see a 
different effect, in which the particles alter the evolution of the Rossby 
vortices and generate a vorticity of their own.
\section{Conclusions}

In this Letter, we have shown that when modelling the self-gravity of the gas and 
solids in protoplanetary disks, gravitational collapse of the solids into 
Mars sized protoplanetary embryos occurs rapidly at the borders of the dead 
zone, where particles concentrate. We have also found that tides from the 
dense gaseous vortices may hinder the formation process significantly

Studies considering the origin of oligarchs 
usually begin from a collection of 10-20 Mars sized objects (e.g., 
Kobuko et al. 2006). This Letter presents the first simulation in which a 
similar number of Mars-sized embryos are formed from centimeter and 
meter sized sized building blocks.

It is crucial to the model that particles grow to sufficient size, 
otherwise the drag force from the gas becomes too strong to allow any
concentration. Testi et al.\ (2003) observed grains of up to cm sizes in
the disk surrounding the pre-main-sequence star CQ Tauri, which provides some
observational evidence that a sufficient number of particles of the required 
size may exist in true protoplanetary disks.

We emphasize again that the model used in this Letter is simplistic, and 
the conditions may differ with a more realistic treatment of the 
dead zone. Nevertheless, the mechanism presented in this Letter (as 
proposed originally by Varni\`ere \& Tagger 2006) appears robust. It 
only requires the RWI to be excited in the borders between the active 
and dead zones, which in turn relies only on the slowdown of the accretion 
flow at this same border. Future studies should include a 
coagulation/fragmentation model such as those of Brauer et al. (2008) or 
Johansen et al. (2008), and focus on the precise state 
of flow at this transition region in global simulations, to address the 
question of how the RWI interacts with the MRI and non-ideal MHD in three 
dimensions.

\begin{acknowledgements}
Simulations were performed at the PIA cluster of the Max-Planck-Institut
f{\"u}r Astronomie and on the Uppsala Multidisciplinary Center for Advanced
Computational Science (UPPMAX). We thank the referee, Dr. Laure Fouchet, 
for many helpful suggestions for improvement of the paper.
\end{acknowledgements}

\Online

\begin{appendix}
\section{Dynamical Equations}

As stated in the main paper, we work in the thin disk approximation, using the vertically integrated
equations of hydrodynamics

\begin{eqnarray}
\ptderiv{\varSigma_g} &=& -\left(\v{u}\cdot\del\right)\varSigma_g -\varSigma_g{\Div\v{u}} + f_D(\varSigma_g) \label{eq:continuity}\\
\ptderiv{\v{u}} &=& -\left(\v{u}\cdot\del\right)\v{u} -\frac{1}{\varSigma_g}\del{P} - \del\varPhi - \frac{\varSigma_p}{\varSigma_g}\v{f}_d \nonumber\\ 
&&+ 2\,\varSigma_g^{-1}\,\Div{\left(\nu\varSigma_g \vt{S}\right)}+ f_\nu(\v{u},\varSigma_g)\label{eq:Navier-Stokes}\\
\frac{d{\v{x}}_{p}}{dt} &=& \v{v}_p\label{eq:particle-vel}\\
\frac{d{\v{v}}_{p}}{dt} &=& - \del\varPhi + \v{f}_d \label{eq:particle}\\
\varPhi&=&\varPhi_{\rm sg} -\frac{GM_\Sun}{r}  \label{eq:potential}\\
\Laplace\varPhi_{\rm sg} &=& 4{\pi}G\left(\varSigma_g+\varSigma_p\right)\delta(z) \label{eq:poisson}\\
P&=&\varSigma_g c_s^2    \label{eq:state}\\
f_d &=& - \left(\frac{3\rho_g C_D |\dv|}{8 a_\bullet \rho_\bullet}\right)\dv.   \label{eq:drag-acelleration}
\end{eqnarray}

In the above equations, $G$ is the gravitational constant, $\varSigma_g$ and $\varSigma_p$ are 
the vertically integrated gas density and bulk density of solids, respectively, 
$\v{u}$ represents the velocity of the gas parcels, $\v{x}_p$ is the position 
and $\v{v}_p$ is the velocity of the solid particles, $P$ is the vertically
integrated pressure, $c_s$ is the sound speed, $\varPhi$ is the gravitational
potential, $\nu$ is the viscosity, and $\vt{S}$ is the rate of strain tensor. The functions 
$f_D(\varSigma_g)$ and $f_\nu(\v{u},\varSigma_g)$ are sixth order hyperdiffusion and 
hyperviscosity terms that provide extra dissipation close to the grid scale, explained 
in Lyra et al. (2008). They are required because the high order scheme of 
the Pencil code has too little overall numerical dissipation. 

The function $\v{f}_d$ is the drag force by which gas and solids interact.
In \Eq{eq:drag-acelleration}, $\rho_\bullet$ is the internal density of a 
solid particle, $a_\bullet$ its radius, and $\dv=\v{v}_p-\v{u}$ its 
velocity relative to the gas. $C_D$ is a dimensionless coefficient that 
defines the strength of the drag force.\\

\section{Drag Force}

The drag regimes are controlled by the mean free path $\lambda$ of 
the gas, which can be expressed in terms of the Knudsen number of the 
flow past the particle $\Kn=\lambda/2a_\bullet$. High Knudsen numbers 
correspond to 
free molecular flow, or Epstein regime. Stokes drag is applicable to low Knudsen numbers. We use the formula of
Woitke \& Helling (2003; see also Paardekooper 2007), which interpolates between Epstein and Stokes regimes
\begin{equation}
  \label{eq:coeff-general}
  C_D = \frac{9\Kn^2 C_D^\Eps + C_D^\Stk}{(3\Kn+1)^2}.
\end{equation}where $C_D^\Eps$ and $C_D^\Stk$ are the coefficients of 
Epstein and Stokes drag, respectively. These coefficients are 
\begin{eqnarray}
  C_D^\Eps &\approx& 2\left(1+\frac{128}{9\pi \Ma^2}\right)^{1/2} \label{eq:coeff-epstein}\\
\vspace{\smallskipamount}\nonumber\\ 
  C_D^\Stk&=&\left\{ \begin{array}{ll}
    24\,\Rey^{-1} + 3.6\,\Rey^{-0.313}  & \mbox{; $\Rey \leq 500$};\\
    \xtimes{9.5}{-5}\,\Rey^{1.397} & \mbox{; $500 < \Rey \leq 1500$};\\
    2.61  & \mbox{; $\Rey > 1500$,} \end{array} \right. \label{eq:coeff-stokes}
\end{eqnarray} where $\Ma=|\dv|/c_s$ is the Mach number, $\Rey= 2a_\bullet\rho |\dv|/\mu$ 
is the Reynolds number of the flow past the particle, and $\mu=\sqrt{8/\pi}\rho c_s \lambda/3$ 
 is the kinematic viscosity of the gas.

The approximation for 
Epstein drag (Kwok 1975) connects regimes of low and high Mach number 
with a good degree of accuracy, and is more numerically friendly than the general case 
(Baines et al. 1965). The piecewise function for the Stokes regime are empirical corrections to 
Stokes law ($C_D=24\Rey^{-1}$), which only applies for low Reynolds numbers.\\

\section{Gas tides and mass loss}

The most remarkable feature of Fig.~3a of the main paper is the mass loss event at 90 orbits. 
It consists of the detachment of a 0.8 $M_{\rm Mars}$ object from the original 
cluster, of 6.7$M_{\rm Mars}$.

\begin{figure*}
  \begin{center}
    \includegraphics[width=\textwidth]{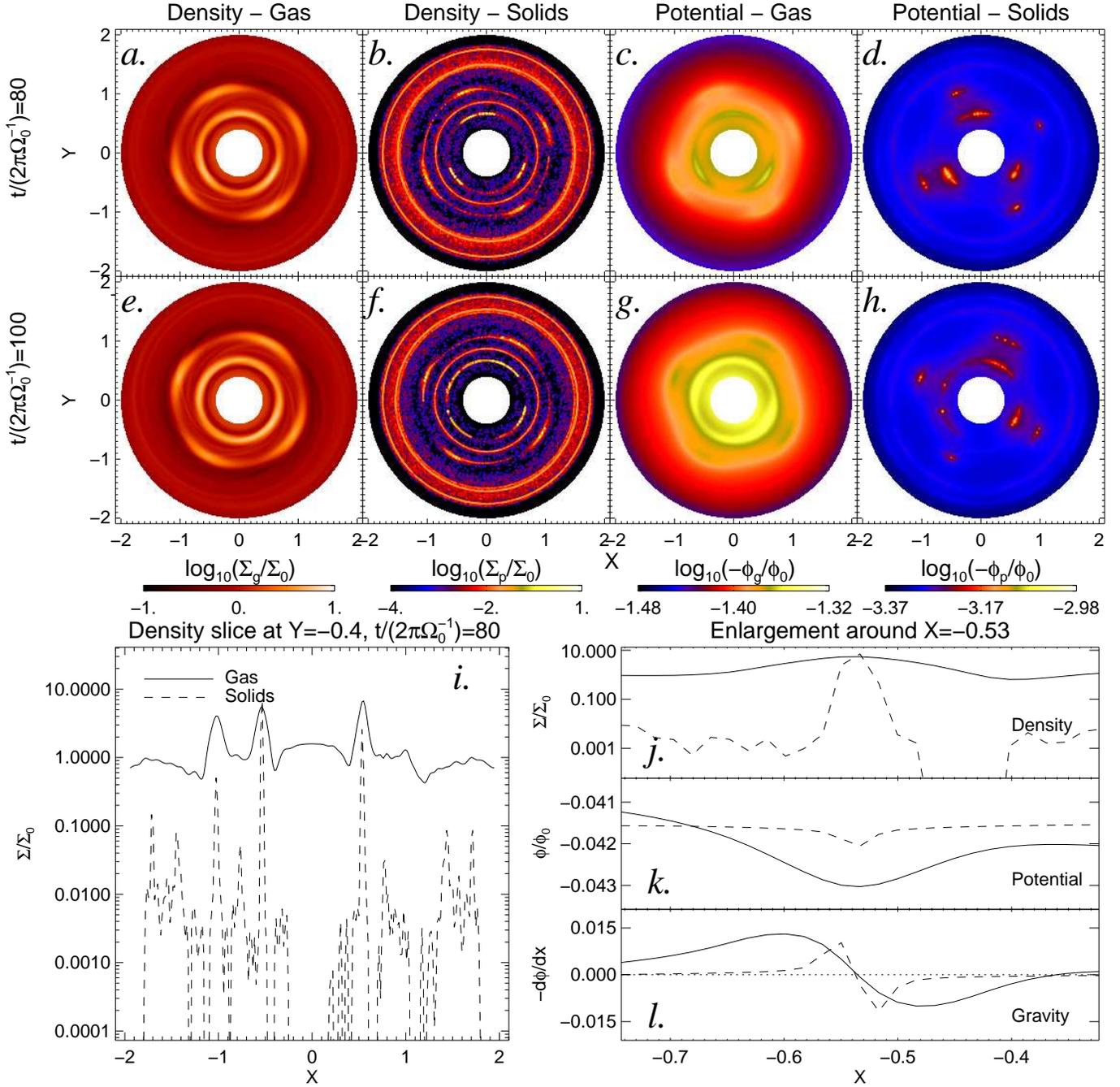}
  \end{center}
  \caption[]{The state of the disk before ($a$-$d$) and after ($e$-$h$) 
the mass-loss episode. The conspicuous differences between them are due to 
the inner vortices passing from the $m$=3 to the $m$=2 mode. As a result, the 
the most massive embryo has left its parental vortex between one snapshot 
and another. It is seen as a bright spot in panels $f$ and $h$, 
at (X,Y)=(-0.65,-0.19). In panel $b$ (before the mass loss), the embryo 
is at (X,Y)=(-0.40,-0.53) but not easily identified among the swarm of solids 
inside the vortex. Panel $i$ shows a horizontal slice through this 
location, in which we see that the peak in the density of solids is not much 
higher than the gas density at the location of the embryo (panel $j$). 
Significant gas tides are expected because the gravitational potential 
(panel $k$) and acceleration (panel $l$) have similar contributions 
from gas and solid components.}
  \label{fig:tides}
\end{figure*}

We observe evidence that this puzzling behaviour is due to gravitational tides from the gas. 
The gas is too pressure-supported to undergo collapse, but the vortices concentrate 
a sufficient amount of material to exert a considerable gravitational pull. We illustrate this 
in \fig{fig:tides}, where we show the state of the disk before (at 80 orbits, \fig{fig:tides}a-\fig{fig:tides}d) the mass-loss episode and after (at 100 orbits, 
\fig{fig:tides}e-\fig{fig:tides}h). 
The plots show the surface densities of gas and solids, and the potential associated 
with them. Even though the clumping of solids yield a considerable gravitational 
pull (\fig{fig:tides}d and \fig{fig:tides}h), these figures indicate that the dominant 
contribution to the gravitational potential comes from the gas - more specifically 
from the vortices, where the gas density peaks at a value one order of magnitude higher than that of the initial conditions. 

The most massive clump is located in the inner disk at 
(X,Y)=(-0.40,-0.53) in \fig{fig:tides}b, not clearly identifiable 
amidst the other particles trapped inside the vortex. However, the embryo is 
immediately observable as the bright point at (X,Y)=(-0.65,-0.19) in \fig{fig:tides}h 
(also visible in \fig{fig:tides}f, albeit less prominently). There 
are two features in this plot that are worth noting. 
First, by comparing the location of the embryo in these figures with the location of the 
vortices, we notice that the planet has left its parental vortex. Second, the inner 
vortices have undergone the transition from the $m$=3 to the $m$=2 mode. Due to 
merging, their gas density has increased, with dramatic consequences for the embryos within 
them. 

We assess how the gravity of the gas influences the motion of the particles 
in \fig{fig:tides}i-\fig{fig:tides}l. In \fig{fig:tides}i, we take a horizontal 
density slice at the position 
of the most massive embryo at 80 orbits. \Figure{fig:tides}j is an enlargement of 
\fig{fig:tides}i about X=-0.53, where the embryo is located. We see that the 
densities of solids and gas peak at similar values. The subsequent figures show the gravitational 
potential (\fig{fig:tides}k) and acceleration (\fig{fig:tides}l) about the embryo. 
The gas produces a deeper gravitational well, albeit smoother than the one displayed 
by the solids. In the acceleration plot, it is seen that the pull of the gas is 
more significant 
than the pull of the embryo at a distance of only 0.26 AU (0.03 in code units, 
corresponding to two grid cells) away from the center. And even where the pull of 
the solids is strongest (one grid cell away from the center), the gravity of 
the gas is still an appreciable fraction of the gravity of the solids. Tides from 
the gas are unavoidable. 

It is beyond the scope of this paper to consider the full mathematical details 
of the theory of tides, especially because the two bodies (the vortex and the 
embryo) are extended. Instead, we consider the following toy model. The tidal force 
$F_T$ experienced by the planet is proportional to the gradient of the acceleration $a$
induced by the gas. It is also proportional to the radius $R$ of the planet: 
$F_T \propto R\,\grad{a}$. Since $\grad{a}=-\Laplace{\varPhi}$, according to 
the Poisson equation, the tidal force is proportional to the local value of the density

\begin{equation}
  F_T \propto R\,\rho_g . 
\end{equation}

We consider the 3D volume density to avoid the requirement of using the Dirac 
delta in the 2D case. Considering 
the planet spherical, Newton's second theorem holds and we can assume that $F_G= -GM/R^2$
for the planet's (self-)gravitational force at its surface. Substituting $M =4/3 \pi \rho_p R^3$, 
we have $F_G\propto R \rho_p$, so 

\begin{equation}
  \zeta=\frac{F_T}{F_G} \propto \frac{\rho_g}{\rho_p},
  \label{tide-to-selfgrav}
\end{equation}i.e., the ratio of the disrupting 
tidal stresses to the self-gravitating forces that attempt
to keep the planet together is directly proportional to the gas-to-solids 
ratio. At 80 orbits, as seen in \fig{fig:tides}j, this ratio is around unity. 
As the vortices undergo the transition from the $m$=3 to the $m$=2 mode, their peak density 
increases (while the planet remains at constant mass). The tides 
eventually become sufficiently strong for a mass-loss event to occur. We also witness some 
of the other, less massive, embryos being completely obliterated by the 
gas tides. Erosion is also important, since we observe a size dependency in this 
effect, with embryos consisting of $a_\bullet$=10\,cm particles being more prone 
to disruption than those consisting of $a_\bullet$=30\,cm particles. We performed 
tests that indicated that the erosion of bound clumps by ram pressure seen by 
Cuzzi et al. (2008) only occurs for clumps consisting of particles smaller than 
cm-size 
for our initial nebula parameters, larger particles being more stable. However, when 
the gas density of the vortices reaches a maximum value a factor ten higher than the initial density, 
the gas drag also strengthens, shifting the onset of erosion towards larger particle 
radii.

The effect of tides will probably be less dramatic in 3D simulations because, as 
the particles settle into the midplane, 
the ratio of the volume gas density to the bulk density of solids 
$\rho_g$/$\rho_p$ is expected to be much lower than the ratio of the 
column gas density to the vertically integrated surface density of solids 
$\varSigma_g/\varSigma_p$.

\end{appendix}

\end{document}